\documentclass[nojss]{jss}

\usepackage{orcidlink,thumbpdf,lmodern,dsfont}
\usepackage{amsmath}

\newcommand{\fct}[1]{\code{#1()}}
\newcommand{\LMjl}{\pkg{LongMemory.jl}}
\newcommand{\Julia}{\proglang{Julia}}
\newcommand{\R}{\proglang{R}}

\author{J. Eduardo Vera-Valdés~\orcidlink{0000-0002-0337-8055}\\Aalborg University}
\Plainauthor{J. Eduardo Vera-Valdés}

\title{\pkg{LongMemory.jl}: Generating, Estimating, and Forecasting Long Memory Models in \proglang{Julia}}
\Plaintitle{LongMemory.jl: Generating, Estimating, and Forecasting Long Memory Models in Julia}
\Shorttitle{\pkg{LongMemory.jl}: Long Memory Modelling in \proglang{Julia}}

\Abstract{
  \pkg{LongMemory.jl} is a package for time series long memory modelling in \Julia\ \citep{juliaprimer}. The package provides functions to generate long memory, estimate model parameters, and forecast. Generating methods include fractional differencing, stochastic error duration, and cross-sectional aggregation. Estimators include the classic ones used to estimate the Hurst effect, those inspired by log-periodogram regression, and parametric ones. Forecasting is provided for all parametric estimators. Moreover, the package adds plotting capabilities to illustrate long memory dynamics and forecasting. This article presents the theoretical developments for long memory modelling, show examples using the data included with the package, and compares the properties of \LMjl\ with current alternatives, including benchmarks. For some of the theoretical developments, \LMjl\ provides the first publicly available implementation in any programming language. A notable feature of this package is that all functions are implemented in the same programming language, taking advantage of the ease of use and speed provided by \Julia. Therefore, all code is accessible to the user. Multiple dispatch, a novel feature of the language, is used to speed computations and provide consistent calls to related methods. The package is related to the \proglang{R} \citep{R:lang} packages \pkg{LongMemoryTS} \citep{R:LongMemoryTS} and \pkg{fracdiff} \citep{R:fracdiff}. 
}

\Plainkeywords{Long memory, long-range dependence, fractional difference, ARFIMA, strong persistence, Julia}

\Address{
  J. Eduardo Vera-Valdés\\
  Department of Mathematical Sciences\\
  Aalborg University\\
  Skjernvej 4A, 9220 Aalborg East, Denmark\\
  Email: \email{eduardo@math.aau.dk}\\
  URL: \url{https://everval.github.io/}
}

\begin{document}


\section[Introduction]{Introduction}\label{sec:intro}


Long memory in time series analysis deals with the notion that data may have a strong dependence on past values. \cite{Hurst1956} is one of the pioneering works on long memory. The author analysed the flow of the Nile River and noted that water reservoirs that do not account for its long-term dynamics are at risk of overflowing. Long memory models are used in finance, biology, economics, climate, and many other fields \citep[see][for several examples]{Beran1994,Palma2006,Beran2013,VeraValdes2021b}.

We say that a time series $x_t$ has long memory if:
\begin{equation}\label{def:cov}
    \gamma_x(k) \approx C_x k^{2d-1}\quad \textnormal{as}\quad k\to\infty,
\end{equation}
where $\gamma_x(k)$ is the autocovariance function and $C_x$ a constant, or if:
\begin{equation}\label{def:spectral}
    f_x(\lambda)\approx C_f\lambda^{-2d}\quad \textnormal{as}\quad \lambda\to 0,
\end{equation}
where $f_x(\lambda)$ is the spectral density function and $C_f$ is a constant. Above, $g(x)\approx h(x)$ as $x\to x_0$ means that $g(x)/h(x)$ converges to $1$ as $x$ tends to $x_0$.

Properties (\ref{def:cov}) and (\ref{def:spectral}) can be analysed graphically by plotting the autocorrelation and periodogram (an estimator of the spectral density), respectively. \LMjl\ provides the functions \fct{autocorrelation\_plot} and \fct{periodogram\_plot} to generate these plots.

\begin{figure}[ht!]
    \centering
    \includegraphics[width=0.9\columnwidth]{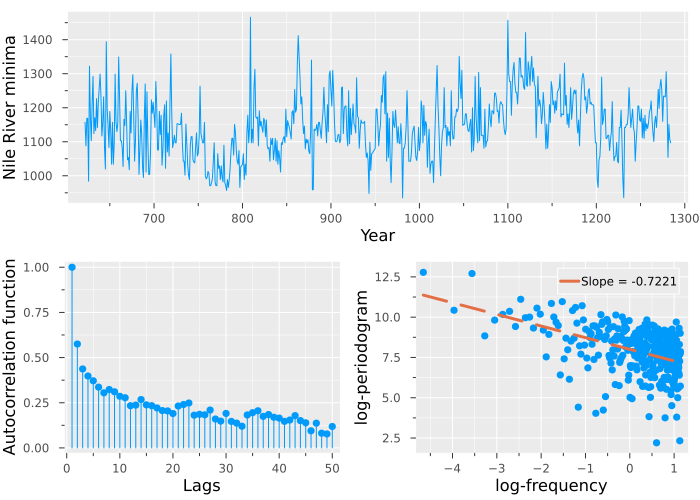}
    \caption[]{Nile River minima (top), its autocorrelation function (bottom left), and log-periodogram (bottom right).}\label{fig:Nile}
\end{figure}

As an example, Figure~\ref{fig:Nile} shows the autocorrelation and periodogram (in logs) for the Nile River minima data. The data are available in \pkg{LongMemory.jl} through \fct{NileData}. The function returns a data frame with columns for year and Nile minima retrieved as \code{NileData().Year} and \code{NileData().NileMin}, respectively. The commands to obtain the figure are shown below.

As the figure shows, the autocorrelation function decays slowly and the periodogram diverges towards infinity near the origin. These are the features of long memory processes described in (\ref{def:cov}) and (\ref{def:spectral}), respectively. This article and the \pkg{LongMemory.jl} package are concerned with methods for modelling data with this type of behaviour.

The following code is used to produce Figure~\ref{fig:Nile}. The data are loaded using the function \fct{NileData} and plotted. The code then plots the autocorrelation function using the function \fct{autocorrelation\_plot} and the log-periodogram using the function \fct{periodogram\_plot}.
\begin{CodeChunk}
    \begin{CodeInput}
    using LongMemory, DataFrames, CSV

    Data = NileData()
    p1 = plot( Data.Year , Data.NileMin, xlabel="Year" ,
        ylabel = "Nile River minima" , legend = false )
    p2 = autocorrelation_plot( Data.NileMin , 50 )
    p3 = periodogram_plot( Data.NileMin )
    l = @layout [a; b c]
    theme(:ggplot2)
    plot(p1, p2, p3, layout = l, size = (700, 500) )
    \end{CodeInput}
\end{CodeChunk}

Note that the figure above can be succinctly obtained, together with the log-variance plot discussed in Section~\ref{sec:variance_plot}, with the function \fct{NileDataPlot} using \LMjl.

The remainder of this article is structured as follows. Sections \ref{sec:generation} to \ref{sec:forecasting} present the theoretical developments behind long memory generation, estimation, and forecasting; respectively. Applications of these developments are shown using \LMjl\ in each of these sections. One of the main advantages of \pkg{LongMemory.jl} is that it is written in \proglang{Julia}, a modern high-performance programming language, as shown by the benchmarks presented in Section \ref{sec:software}. Section \ref{sec:conclusion} concludes. Notebooks with all the code used in this article and additional examples are available on the package's website at \href{https://everval.github.io/LongMemory.jl/}{everval.github.io/LongMemory.jl/}

\section{Long memory generation} \label{sec:generation}

In the time series literature, several models have been proposed to generate long memory. In this section, we present the most widely used models in applied work.

\subsection{Fractional differencing}

The notion of slowly decaying autocorrelations shown in $(\ref{def:cov})$ was one of the motivations behind the fractional difference operator \citep{Granger1980b,Hosking1981}. As the name suggests, the fractional difference operator extends the difference operator to fractional values. It is defined as:
\begin{equation}\label{eq:frac_diff}
    x_t = (1-L)^d\varepsilon_t,
\end{equation}
where $L$ is the lag operator, $\varepsilon_t$ is a white noise process with variance $\sigma^2$, and $d\in(-1/2,1/2)$. The fractional difference operator is decomposed using the binomial expansion to generate the series as:
\begin{equation}\label{eq:arfima_ma}
    x_t = \sum_{k=0}^{\infty} \pi_k \varepsilon_{t-k},
\end{equation}
with coefficients $\pi_k=\Gamma(k+d)/(\Gamma(d)\Gamma(k+1))$ for $k\in\mathds{N}$, where $\Gamma()$ is the gamma function.

The autocorrelation function for a fractionally differenced process, $\rho_{I(d)}(k)$, is given by:
\begin{equation}\label{ACF:Id}
    \rho_{I(d)}(k) = \frac{\Gamma(k+d)\Gamma(1-d)}{\Gamma(k-d+1)\Gamma(d)}.
\end{equation}

Asymptotically, Stirling's approximation is used to show that $\rho_{I(d)}(k)\approx k^{2d-1}$ as $k\to\infty$. Hence,  the process observes Equation~\ref{def:cov}. Additional properties of the fractional difference operator have been well documented in, among others, \cite{Baillie1996} and \cite{Beran2013}.

The next lines of code show how to generate a series with long memory using the fractional difference operator and plot it along with its autocorrelation function. We use the functions \fct{fi\_gen} and \fct{autocorrelation} in \pkg{LongMemory.jl} to generate the series and compute its autocorrelation. Note that we fixed a random seed for reproducibility.
\begin{CodeChunk}
    \begin{CodeInput}
    using LongMemory, Plots, Random
    Random.seed!( 1234 )

    dx = fi_ gen( 100 , 0.3 )
    p1 = plot( dx , label = "Fractionally differenced data" )
    p2 = plot( 0:50 , autocorrelation( dx , 51 ) ,
        label = "Sample autocorrelation function", line = :stem ,
        marker = :circle)
    plot!( 0:50, fi_cor_vals( 51, 0.3 ),
    label = "Theoretical autocorrelation function")
    l = @layout [a b]
    theme(:ggplot2)
    plot(p1, p2, layout = l, size = (700, 300) )
    \end{CodeInput}
\end{CodeChunk}

\begin{figure}[ht!]
    \centering
    \includegraphics[width=0.9\columnwidth]{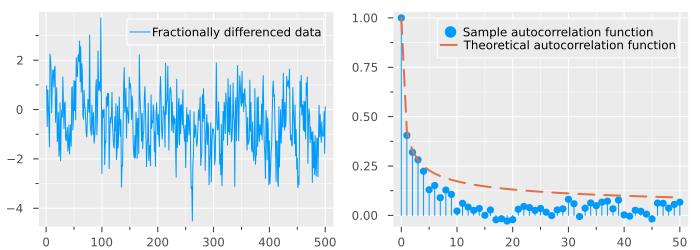}
    \caption[]{Fractionally differenced data (left), and its sample autocorrelation function together with the theoretical one (right).}\label{fig:fracdiff}
\end{figure}

The output of the code is shown in Figure~\ref{fig:fracdiff}. Note that the figure also shows the theoretical autocorrelation function for the fractional difference operator obtained using the function \fct{fi\_cor\_vals} in \pkg{LongMemory.jl}. The theoretical autocovariance is available using the function \fct{fi\_var\_vals}.

Under the hood, \pkg{LongMemory.jl} uses the fast algorithm for fractional differencing proposed by \cite{Jensen2014}. The function is available as \fct{fracdiff}, which allows the user to specify the process that should be fractionally differenced. Moreover, the package uses the recursive algorithm to compute the autocovariance function to speed up the computations. Benchmarks against the non-recursive version are shown in Section~\ref{sec:benchmarks}.


\subsection{Cross-sectional aggregation}\label{sec:csa}

One of the most cited reasons behind the presence of long memory in real data is due to cross-sectional aggregation \citep{Granger1980}. The cross-sectional aggregated process is defined as:
\begin{equation}\label{eq:csa_def}
    x_t = \frac{1}{\sqrt{N}}\sum_{i=1}^N x_{i,t},
\end{equation}
where each $x_{i,t}$ is an autoregressive process defined by
\begin{equation}\label{eq:csa_ind_ar}
    x_{i,t} = \alpha_i x_{i,t-1}+\varepsilon_{i,t},
\end{equation}
where $\varepsilon_{i,t}$ is a white noise process with variance $\sigma^2$ for all $i,t$. The autoregressive parameter $\alpha_i^2$ is sampled from a beta distribution, $\mathcal{B}(\alpha; p, q)$, where $p,q>1$, with density given by:
\begin{equation}\label{eq:beta_dist}
    \mathcal{B}(\alpha; p, q) =  \frac{1}{B(p,q)} \alpha^{p-1}(1-\alpha)^{q-1}\ \ \ \ \textnormal{for}\ \ \ \alpha\in(0,1),
\end{equation}
with $B(\cdot,\cdot)$ the beta function.

As $N\to\infty$, the autocorrelations of $x_t$ are given by
\begin{equation}\label{ACF:CSA}
    \rho_{CSA}(k) = \frac{B(p+k/2,q-1)}{B(p,q-1)},
\end{equation}
which can be shown decay at a hyperbolic rate with parameter $d=1-q/2$. Therefore, $x_t$ has long memory in the sense of Equation~\ref{def:cov}. Note that long memory by cross-sectional aggregation is more flexible due to the reliance on both parameters of the beta distribution (\ref{eq:beta_dist}).

Following Equation~\ref{eq:csa_def}, one way to generate long memory processes is to aggregate autoregressive processes. Alternatively, the moving average representation of the limiting process is given by:
\begin{equation}\label{eq:csa_ma}
    x_t = \sum_{k=0}^{\infty} \phi_k \varepsilon_{t-k},
\end{equation}
with coefficients $\phi_k = [B(p+k,q)/B(p,q)]^{1/2}$ for $k\in\mathds{N}$. This suggests another way to generate long memory processes by aggregation using a similar algorithm to the one used for fractional differencing.

\LMjl\ uses multiple dispatch, a novel feature of \proglang{Julia}, to provide a unified interface to generate long memory by cross-sectional aggregation using a finite number of autoregressive processes and the asymptotic version using the fast algorithm proposed by \citet{VeraValdes2021a}. The finite version of the algorithm is used if a finite number of autoregressive processes is provided. Otherwise, the asymptotic version is used. 

Given that the result is asymptotic, long memory is present only for a large number of autoregressive processes. \citet{Haldrup2017} shows that the number of autoregressive processes should increase at the same rate as the sample size to obtain a good approximation. Therefore, the asymptotic version is recommended for large sample sizes, as shown in the benchmarks in Section \ref{sec:benchmarks_csa}. The finite version is useful for small sample sizes and to illustrate the dynamics of the process.

Analogous functions to those presented for fractional differencing are provided in \LMjl\ for cross-sectional aggregation. That is, the functions \fct{csa\_gen}, \fct{csa\_var\_vals}, and \fct{csa\_cor\_vals} are available in the package. The functions generate the series and obtain its theoretical autocovariance and autocorrelation, respectively. 

\begin{figure}[ht!]
    \centering
    \includegraphics[width=0.8\columnwidth]{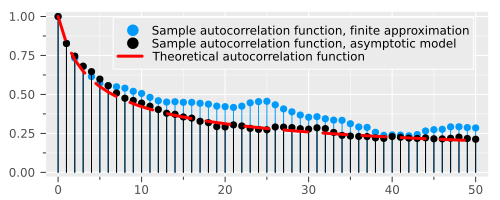}
    \caption[]{Theoretical and sample autocorrelation functions of cross-sectional aggregated data.}\label{fig:csa}
\end{figure}

The following lines of code show how to generate series with long memory using cross-sectional aggregation using the finite version of the algorithm and the asymptotic one. The sample autocorrelations for both series are then plotted together with the theoretical one. The output of the code is shown in Figure~\ref{fig:csa}.
\begin{CodeChunk}
    \begin{CodeInput}
    using LongMemory, Random, Plots

    Random.seed!(1234)
    csa_fin = csa_gen(1000,1000,1.3,1.5)
    csa_asym = csa_gen(1000,1.3,1.5)
    plot( 0:50 , autocorrelation( csa_fin , 51 ) ,
        label = "Sample autocorrelation function, finite approximation",
        line = :stem , marker = :circle )
    plot!( 0:50 , autocorrelation( csa_asym , 51 ) ,
        label = "Sample autocorrelation function, asymptotic model",
        line = :stem , marker = :circle, color = :black )
    plot!( 0:50, csa_cor_vals( 51, 1.3, 1.5 ),
        label = "Theoretical autocorrelation function",
        linewidth = 3, line = :dash, color = :red )
    theme(:ggplot2)
    plot!(size = (500, 200) )
    \end{CodeInput}
\end{CodeChunk}

Note that both series are generated using the function \fct{csa\_gen} using multiple dispatch. The finite version of the algorithm is used by passing the number of autoregressive processes to consider. In the figure, the number of autoregressive processes is equal to the sample size.

\LMjl\ is the first publicly available software in any programming language to generate long memory by cross-sectional aggregation, the most commonly cited reason behind long memory in real data. 


\subsection{Stochastic duration shock}

Another way to generate long memory was proposed by \cite{Parke1999}. The author considered a time series generated as the sum of shocks of stochastic magnitude and duration. If only a small proportion of errors survive for long periods of time, then the resulting series shows long memory according to
Equation~\ref{def:cov}.

The stochastic duration shocks model is defined as:
\begin{equation}\label{eq:edm_def}
    x_t = \sum_{s=-\infty}^t g_{s,t}\varepsilon_s.
\end{equation}
where $\varepsilon_s$ is a white noise process with variance $\sigma^2$ and stochastic duration, where $g_{s,t}$ is the indicator function for the event that the error $\varepsilon_s$ is alive in period $t$. Let $p_k$ be the probability that a shock survives for $k$ periods. If $p_k\approx k^{-2+2d}$ as $k\to\infty$, $x_t$ will have hyperbolic decaying autocorrelations in the sense of Equation~\ref{def:cov}. 


\pkg{LongMemory.jl} deploys the algorithm proposed by \citet{Parke1999} to generate series with stochastic duration shocks through the function \fct{sds\_gen}. Furthermore, the package provides access to the function \fct{fi\_survival\_probs} to compute survival probabilities that mimic the autocovariance of fractional differencing. For the latter, the recursive formulation for generating the series is used to speed up the computations; see Section~\ref{sec:benchmarks_recursive}.

\begin{figure}[ht!]
    \centering
    \includegraphics[width=0.9\columnwidth]{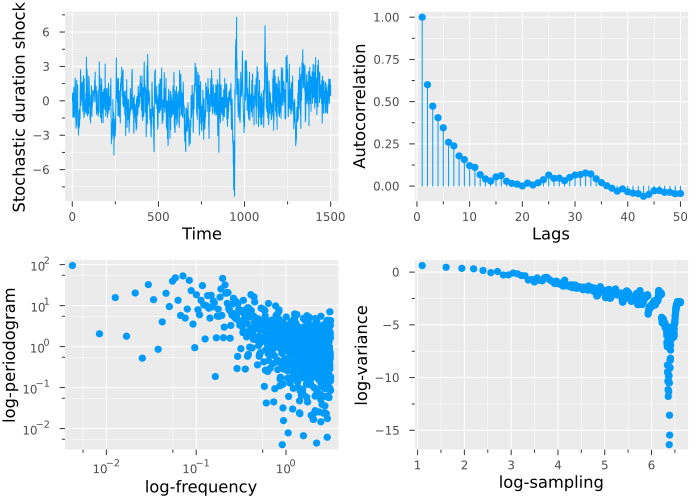}
    \caption[]{Data generated using the stochastic duration shocks model (top left), its sample autocorrelation function (top right), log-periodogram (bottom left), and log-variance plot (bottom right).}\label{fig:edm}
\end{figure}

The following line of code shows how to generate a series with long memory using the stochastic duration shocks model and plot it along with several diagnostics plots for long memory. We use the function \fct{LMPlot} from \pkg{LongMemory.jl} to generate the plot. The output of the code is shown in Figure~\ref{fig:edm}.
\begin{CodeChunk}
    \begin{CodeInput}
    julia> LMPlot(sds_gen(1000,0.45), name = "Stochastic duration shock")
    \end{CodeInput}
\end{CodeChunk}

To the best of our knowledge, this is the first publicly available software to generate long memory using the stochastic duration shock model.


\section{Long memory estimation}\label{sec:estimation}

Several long memory estimators have been proposed in the literature. They range from classics based on the rescaled range statistic used by \citet{Hurst1956}, to ones based on the spectral density property described in Equation~\ref{def:spectral}, and parametric estimators.

\subsection{Classic estimators}

\LMjl\ includes the log-variance estimator and the rescaled range statistic. The following subsections describe each of them. Although they are intuitive, they are not recommended for reliable statistical inference. However, both methods, and particularly the plots associated to them, may serve to assess whether the data may present long memory dynamics or not. Once detected, better statistical models, such as the ones presented in the following sections, should be used for inference.

\subsubsection{Log-variance estimator}\label{sec:variance_plot}

The log-variance plot is a simple method of assessing the presence of long memory in a time series. Note that as a consequence of Equation~\ref{def:cov}, the variance of the sample mean of a time series with long memory follows:
\begin{equation}\label{eq:variance}
    Var\left(\bar{x}\right) = Var\left( \frac{1}{n} \sum_{t=1}^n x_t \right) \approx C_v n^{2d-1},
\end{equation}
where $C_v$ is a constant.

Therefore, one way to determine the presence of long memory is to plot the log of the sample variance of the series against the log of the sample size. If the series has long memory, the plot should be a straight line with slope $2d-1$. In contrast, the variance of the sample mean decreases at a rate of $n^{-1}$ for a short memory process, which implies that the log-variance plot should be a straight line with slope $-1$.

The functions \fct{log\_variance\_est} and \fct{log\_variance\_plot} in \pkg{LongMemory.jl} compute the log-variance estimator and produce the plot, respectively. Moreover, the latter function can optionally display the estimated slope along with the theoretical one for a short memory process for comparison.

\begin{figure}[ht!]
    \centering
    \includegraphics[width=0.65\columnwidth]{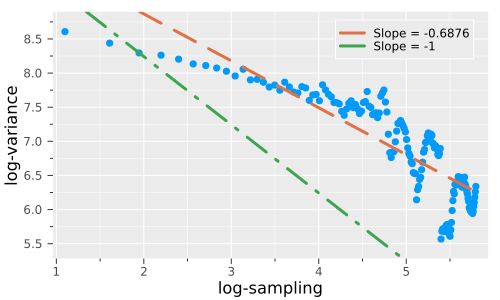}
    \caption[]{Variance plot for the Nile River data.}\label{fig:Nilevariance}
\end{figure}

As illustration, Figure~\ref{fig:Nilevariance} presents the log-variance plot for the Nile River minima data along with the estimated line and the theoretical line of a short memory process. The figure shows that the log-variance is consistent with the data possessing long memory.

The following line of code shows how to generate the plot in Figure~\ref{fig:Nilevariance} using \pkg{LongMemory.jl}, where the Nile River data is loaded using the function \fct{NileData}.
\begin{CodeChunk}
    \begin{CodeInput}
    julia> log_variance_plot( NileData().NileMin; m=300,
            slope = true, slope2 = true )
    \end{CodeInput}
\end{CodeChunk}

Above, the parameter \code{m} is the number of subsamples to use in computing the log-variance, which translates into the number of points in the plot. Parameters \code{slope} and \code{slope2} are used to display the estimated slope and the theoretical one for a short memory process, respectively.

\subsubsection{Rescaled range analysis}\label{sec:R/S}

The rescaled range is the measure calculated by \cite{Hurst1956} to determine the long-term variability in the Nile River data. 

The series range is defined as the difference between the maximum and minimum values of the cumulative series of deviation. That is, the range is given by:
\begin{equation}\label{eq:range}
    R_n(k) = \max_{1\leq k \leq n} \sum_{i=1}^k (x_i-\bar{x})-\min_{1\leq k \leq n} \sum_{i=1}^k (x_i-\bar{x}),
\end{equation}
where $\bar{x}=\frac{1}{n}\sum_{i=1}^{n} x_i$.

The range is then rescaled by the sample standard deviation to obtain the rescaled range:
\begin{equation}\label{eq:R/S}
    R/S = \frac{R_n(k)}{S_n(k)},
\end{equation}
where $S_n^2(k) = \frac{1}{n-1}=\sum_{i=1}^k(x_i-\bar{x})^2$ is the sample variance.

The rescaled range is calculated for different values of $k$, and the results are plotted on a log-log scale. If the series has long memory, the plot should be a straight line with slope $d=H-1/2$, where $H$ is the Hurst exponent.

This suggests that another way to estimate the long memory parameter is by the slope of the rescaled range plot. The function \fct{rescaled\_range\_est} in \pkg{LongMemory.jl} computes the rescaled range estimator, while the function \fct{rescaled\_range\_plot} produces the plot. Moreover, analogous to the log-variance function above, the plotting function can optionally return the estimated slope along with the theoretical one for a short memory process.

As an illustration, the following line of code shows how to estimate the long memory parameter in the Nile River minima time series through the rescaled range using 300 values for $k$ in (\ref{eq:range}). The output of the code is shown below.
\begin{CodeChunk}
    \begin{CodeInput}
    julia> rescaled_range_est( NileData().NileMin; k = 300)
    \end{CodeInput}
    \begin{CodeOutput}
    0.4254606013817649
    \end{CodeOutput}
\end{CodeChunk}

\subsection{Semiparametric estimators}\label{sec:semiparametric}

Semiparametric estimators in the frequency domain are based on Equation~\ref{def:spectral}; that is, the behaviour of the spectrum near the origin. The idea is to evaluate the periodogram of the time series in a vicinity of the origin where the long memory parameter drives the spectral density.

The periodogram is defined as:
\begin{equation}\label{eq:periodogram}
    I_x(\lambda) = \frac{1}{2\pi}\left|\sum_{t=1}^n x_t e^{-i\lambda t}\right|^2,
\end{equation}
where $\lambda\in[0,\pi]$ is the frequency. The periodogram is an unbiased estimator of the spectral density, and it is consistent with the spectral density at the Fourier frequencies.

Note that by evaluating near the origin, the semiparametric estimators circumvent the need to specify the short-term dynamics of the time series. Hence, the semiparametric estimators are robust to short-run dynamics such as observational noise.

The next subsections describe the semiparametric estimators for long memory considered in \LMjl.

\subsubsection{Log-periodogram regression or Geweke-Porter-Hudak estimator}\label{sec:GPH}

\cite{Geweke1983} proposed an estimation procedure based on the log-periodogram regression near the zero frequency. The log-periodogram regression is given by:

\begin{equation}\label{eq:log-periodogram}
    \log(I(\lambda_k)) = c-2d \log(\lambda_k)+u_k,\quad k = 1,\cdots,m,
\end{equation}
where $I(\lambda_k)$ is the periodogram of $x_t$, $\lambda_{k} = e^{i2\pi k /T}$ are the Fourier frequencies, $c$ is a constant, $u_k$ is the error term, and $m$ is a bandwidth parameter that grows with the sample size.

The consistency and asymptotic normality of the log-periodogram regression were proven by \cite{Robinson1995a}. Denote $\hat{d}_{GPH}$ to the estimate of the long memory parameter via the log-periodogram regression, then:
\begin{equation}\label{Eq:LPR_Dist}
    \sqrt{m}(\hat{d}_{GPH}-d) \xrightarrow[d]{} N\left(0,\frac{\pi^2}{24} \right),
\end{equation}
where $m$ is the bandwidth as before, and $\xrightarrow[d]{}$ denotes convergence in distribution.

An example showing how to estimate the long memory parameter in the Nile River minima time series using the log-periodogram regression is presented below along with the discussion on the bias-reduced version.

\subsubsection{Bias-reduced log-periodogram regression}\label{sec:AG}

\cite{Andrews2003} proposed to replace the constant in (\ref{eq:log-periodogram}) with a polynomial in $\lambda_{k}^2$ to reduce the bias. In the typical bias-variance trade-off, the reduction in bias comes at the cost of an increase in the variance, which depends on the degree of the polynomial used for estimation.

\LMjl\ allows the practitioner to determine the number of polynomial terms to add, recovering the GPH estimator if no bias reduction terms are added (the default). To our knowledge, \LMjl\ is the only software that allows the user to specify the number of polynomial terms to add to the log-periodogram regression.

The following code shows the use of the function \fct{gph\_est} to estimate the long memory parameter in the Nile River minima time series using the GPH and bias-reduced log-periodogram regression with one polynomial term by setting \code{br = 1}.
\begin{CodeChunk}
    \begin{CodeInput}
    julia> (gph_est( NileData().NileMin ),
            gph_est( NileData().NileMin ; br = 1))
    \end{CodeInput}
    \begin{CodeOutput}
    (0.37449410505423664, 0.39745526593583125)
    \end{CodeOutput}
\end{CodeChunk}

The above results are presented using a tuple in \Julia. The first element is the GPH estimator, and the second is the bias-reduced log-periodogram regression with one polynomial term. The bandwidths are left at the default value of $m=T^{4/5}$, where $T$ is the sample size.

The variance of the estimators is calculated with the function \fct{gph\_est\_variance} as shown in the code below.
\begin{CodeChunk}
    \begin{CodeInput}
    julia> (gph_est_variance( NileData().NileMin ),
        gph_est_variance( NileData().NileMin ; br = 1))
    \end{CodeInput}
    \begin{CodeOutput}
    (0.002272008379624622, 0.0051120188541553995)
    \end{CodeOutput}
\end{CodeChunk}

The variance inflating factors were calculated by \cite{Andrews2003} and are automatically used by \LMjl. Above, note that the variance is $2.25$ times larger for the bias-reduced log-periodogram regression with one polynomial term than for the GPH estimator.

As shown in (\ref{Eq:LPR_Dist}), the variance depends only on the bandwidth parameter $m$, which in turn depends only on the sample size. Therefore, only these two parameters are needed to obtain the variance. \LMjl\ makes use of the multiple dispatch implemented in \Julia\ to compute the variance of the estimators using only the sample size and bandwidth. Therefore, the above variances can alternatively be computed with the following code.
\begin{CodeChunk}
    \begin{CodeInput}
    julia> (gph_est_variance( length( NileData().NileMin ) ),
        gph_est_variance( length( NileData().NileMin ) ; br = 1))
    \end{CodeInput}
    \begin{CodeOutput}
    (0.002272008379624622, 0.0051120188541553995)
    \end{CodeOutput}
\end{CodeChunk}

Above, we use the \fct{length} function to obtain the sample size, and the bandwidth is left at the default value of $m=T^{4/5}$, where $T$ is the sample size.



\subsubsection{Local Whittle estimator}\label{sec:whittle}

An alternative semiparametric formulation was developed by \cite{Kunsch1987}. The author proposed to estimate the parameter as the minimiser of the local Whittle likelihood function given by
\begin{equation}\label{eq:LW}
    R(d) = \log\left(\frac{1}{m}\sum_{k=1}^{m}\lambda_k^{2d}I(\lambda_k)\right)-\frac{2d}{m}\sum_{k=1}^{m}\log(\lambda_k),
\end{equation}
where $I(\lambda_k)$ is the periodogram of $x_t$, $\lambda_{k} = e^{i2\pi k /T}$ are the Fourier frequencies, and $m$ is the bandwidth parameter.

In contrast to log-periodogram regression, the local Whittle estimator requires numerical optimisation for estimation. \LMjl\ uses \pkg{Optim.jl} \citep{Mogensen2018} to minimise the local Whittle likelihood function. 

\cite{Robinson1995} proved the consistency and asymptotic normality of the local Whittle estimator. Denote $\hat{d}_{LW}$ to the estimate of the long memory parameter via the local Whittle estimator, then:
\begin{equation}\label{Eq:LW_Dist}
    \sqrt{m}(\hat{d}_{LW}-d) \xrightarrow[d]{} N\left(0,\frac{1}{4} \right),
\end{equation}
where $m$ is the bandwidth as before.

An example showing how to estimate the long memory parameter in the Nile River minima time series using the local Whittle approach is presented below along with the discussion on its refinement.

\subsubsection{Exact local Whittle estimator}\label{sec:exact_whittle}

A refinement of the local Whittle approach was suggested by \cite{Shimotsu2005}. The author proposed the exact local Whittle estimator as the minimiser of the function given by:
\begin{equation}
R(d) = \log\left(\frac{1}{m}\sum_{k=1}^{m}I_{\Delta^d}(\lambda_k)\right)-\frac{2d}{m}\sum_{k=1}^{m}\log(\lambda_k),
\end{equation}
where $I_{\Delta^d}(\lambda_k)$ is the periodogram of $(1-L)^d x_t$, where $(1-L)^d$ is the fractional difference operator as before, $\lambda_{k} = e^{i2\pi k /T}$ are the Fourier frequencies, and $m$ is the bandwidth parameter.

Note that the consistency and asymptotic normality of the ELW are the same as those for the local Whittle estimator. However, the exact local Whittle estimator is restricted to zero-mean processes. \LMjl\ deploys the feasible version of the estimator by demeaning the data first.

The following lines of code show how to estimate the long memory parameter in the Nile River minima time series using the local Whittle and exact local Whittle estimators in \LMjl. The bandwidth is left at the default value of $m=T^{4/5}$, where $T$ is the sample size.
\begin{CodeChunk}
    \begin{CodeInput}
    julia> (whittle_est( NileData().NileMin ),
        exact_whittle_est( NileData().NileMin ))
    \end{CodeInput}
    \begin{CodeOutput}
    (0.37635955766433826, 0.4088495239569418)
    \end{CodeOutput}
\end{CodeChunk}

Variance estimates for the local Whittle estimators and the exact local Whittle estimators are obtained using the functions \fct{whittle\_est\_variance} and \fct{exact\_whittle\_est\_variance}, respectively. The following line of code obtains the variance estimates for the time series of the Nile River minima.
\begin{CodeChunk}
    \begin{CodeInput}
    julia> (whittle_est_variance( NileData().NileMin ),
        exact_whittle_est_variance( NileData().NileMin ))
    \end{CodeInput}
    \begin{CodeOutput}
    (0.0013812154696132596, 0.0013812154696132596)
    \end{CodeOutput}
\end{CodeChunk}

Note that by construction, both variance estimators are the same, depending only on the bandwidth parameter $m$ and the sample size. Similarly to the log-periodogram case, \LMjl\ uses multiple dispatch to alternatively compute the variance of the estimators using only the sample size and bandwidth.


\subsection{Parametric estimators}\label{sec:parametric}

Parametric estimators are based on the autocovariance function of the time series. The idea is to fit a parametric model to the autocovariance function and estimate the long memory parameter from the model. \LMjl\ includes the maximum likelihood estimator (MLE) for fractional differencing, MLE for cross-sectional aggregation, and heterogeneous autoregressive estimator. The following subsections describe each of them.

\subsubsection{MLE for fractional difference}\label{sec:fimle}

Let $X=[x_0,\cdots,x_{T-1}]^\top$ be a sample of size $T$ of a fractionally differenced time series, Equation~\ref{eq:frac_diff}, and let $\theta = [d,\sigma^2]^\top$. Under the assumption that $\varepsilon_t$ follows a normal distribution, $X$ follows a normal distribution with probability density given by:
\begin{equation}
    f(\theta|X) = (2\pi)^{-T/2}|\Sigma|^{-1/2}\exp\left(-\frac{1}{2}X^\top\Sigma^{-1}X\right),
\end{equation}
where $\Sigma$ is the covariance matrix defined as:
\begin{equation}
\Sigma = \sigma^2\rho_{I(d)}(0)\left[\rho_{I(d)}(|j-k|)\right]_{j,k=1},    
\end{equation}
with $\rho_{I(d)}(k)$ the autocorrelation function in (\ref{ACF:Id}).

We estimate the parameters by maximising the log-likelihood:
\begin{equation}\label{ecn:minMLE_fracdiff}
    \hat{\theta} = \max_{\theta} \log(f(\theta|X)).
\end{equation}

The estimator is consistent and normally distributed. Furthermore, under additional regularity conditions, the estimator is asymptotically normally distributed even under misspecification of the error distribution \citep{Sowell1992}.

For implementation purposes, we isolate $\sigma^2$ to reduce computational burden by reducing the number of parameters to estimate \citep{Doornik2003}. Let $\Sigma = \sigma^2\Gamma$, and estimate the long memory parameter by:
\begin{equation}
    \hat{d}_{MLE} = \max_{d}\frac{1}{2T}\log |\Gamma| + \frac{1}{2}\log(T^{-1}X^\top\Gamma^{-1}X),
\end{equation}
where we discarded the constant and divided by $T$ to reduce the effect of the sample size on convergence.

The variance of the error term is then recovered by:
\begin{equation}
    \hat{\sigma}^2 = T^{-1}X^\top\hat{\Gamma}^{-1}X,    
\end{equation}
where we obtain $\hat{\Gamma}$ by substituting $\hat{d}$.

\LMjl\ provides the function \fct{fi\_mle\_est} to estimate the long memory parameter using the maximum likelihood estimator for fractional differencing. The function returns a tuple with the estimated long memory parameter and the estimated variance of the error term.

The next lines of code show how to estimate the long memory parameter and standard deviation of the error term for the Nile River minima time series using the maximum likelihood estimator for fractional differencing.
\begin{CodeChunk}
    \begin{CodeInput}
    julia> fi_mle_est( NileData().NileMin )
    \end{CodeInput}
    \begin{CodeOutput}
    (0.3925714993964694, 69.95632676539786)
    \end{CodeOutput}
\end{CodeChunk}


\subsubsection{MLE for cross-sectional aggregation}\label{sec:csamle}

Maximum likelihood estimation of the cross-sectional aggregated process proceeds in a manner similar to that in the fractional differentiation case. The only difference is that the autocorrelation function is given by (\ref{ACF:CSA}) instead of (\ref{ACF:Id}). However, note that (\ref{ACF:CSA}) depends on two parameters, where the second controls the rate of decay of the autocorrelation function.

The following lines of code show how to estimate the long memory parameter in the Nile River minima time series using the maximum likelihood estimator for cross-sectional aggregation in \LMjl.
\begin{CodeChunk}
    \begin{CodeInput}
    julia> csa_mle_est( NileData().NileMin )
    \end{CodeInput}
    \begin{CodeOutput}
    (1.0000010468704805, 2.447721694890551, 106.79804259351367)
    \end{CodeOutput}
\end{CodeChunk}

The function returns a tuple with the estimated parameters for the beta distribution and standard deviation of the error term. The parameters are returned in the order $(a,b,\sigma^2)$.

\subsubsection{Heterogenous autoregressive estimator}\label{sec:har}

The heterogeneous autoregressive model (HAR) of \cite{Corsi2009} has been used in the time series literature due to its ability to mimic long memory behaviour without being a long memory model. Of particular relevance is the $HAR(3)$ model, a constrained $AR(22)$, given by:
\begin{equation}\label{eqn:har_model}
    x_t = a_0+a_1x_{t-1}^{(f)}+a_2 x_{t-1}^{(w)} + a_3 x_{t-1}^{(m)}+\epsilon_t,
\end{equation}
where $x_{t-1}^{(f)}=x_{t-1}$, $x_{t-1}^{(w)}=\frac{1}{5}\sum_{i=1}^5{x_{t-i}}$ and, $x_{t-1}^{(m)}=\frac{1}{22}\sum_{i=1}^{22}{x_{t-i}}$. In finance, the specification aims to model the behaviour of different agents responding to uncertainty at different horizons. The three components capture the daily $(x_t^{(f)})$, weekly $(x_t^{(w)})$, and monthly $(x_t^{(m)})$ levels of uncertainty. Equation~\ref{eqn:har_model} is estimated by ordinary least squares.

\LMjl\ allows the user to specify the number of lags to use in the HAR model, defaulting to the HAR($3$) model above. The following lines of code show how to estimate the long memory parameter in the Nile River minima time series using a custom HAR($2$) specification containing components for the first and seventh lags. The motivation for the HAR($2$) specification is to model the \textit{seven years of great abundance followed by seven years of famine} that inspired Mandelbrot's Joseph effect \citep{Mandelbrot1968}.
\begin{CodeChunk}
    \begin{CodeInput}
    julia> har_est( NileData().NileMin ; m = [1,7] )
    \end{CodeInput}
    \begin{CodeOutput}
    ([254.23541690816745, 0.40096895301134294, 0.377482428389992],
        69.6124509836161)
    \end{CodeOutput}
\end{CodeChunk}

We specify the lags to include using the optional input array \code{m}. The function returns a tuple whose first entry is an array with estimated parameters of the HAR model. The second element of the tuple presents the estimated standard deviation of the error term. The parameters of the HAR specification are returned in the order $[a_0,a_1,a_2]$.

\section{Long memory forecasting}\label{sec:forecasting}

Long memory forecasting relies on the parametric models described in Section~\ref{sec:parametric}. The idea is to estimate the parameters of the model using the first $T$ observations and then use the model to forecast the next $h$ observations. Forecasting is done by obtaining the autoregressive formulation of the model and using it recursively to generate the next observation. That is, we obtain the parameters in the specification given by:
\begin{equation}\label{eqn:forecasting}
    x_{t} = \sum_{k=-\infty}^{t-1}\psi_k x_{t-k}+\epsilon_{t}.
\end{equation}

In practice, the infinite sum is truncated to a finite number of lags depending on the sample, and the error term is assumed to be zero-mean.

\LMjl\ includes the functions \fct{fi\_forecast}, \fct{csa\_forecast}, and \fct{har\_forecast} to produce forecasts for fractional differencing, cross-sectional aggregation, and heterogeneous autoregressive models. The following subsections describe each of them. Section \ref{sec:forecasting_nile} shows an example using the Nile Rive minima data.

\subsubsection{Forecasting the fractional difference model}\label{sec:forecasting_fracdiff}

The autoregressive coefficients for the fractional difference model are shown to be $\psi_k = \Gamma(k-d)/[\Gamma(-d)\Gamma(k+1)]$, where $\Gamma(\cdot)$ is the gamma function. The function \fct{fi\_ar\_coefs} in \LMjl\ computes them using a recursive implementation to reduce computational time, see Section \ref{sec:benchmarks_recursive}. The forecasts are obtained using the function \fct{fi\_forecast}, while the function \fct{fi\_forecast\_plot} produces the plot. Optionally, the confidence bands for the forecast can be shown by specifying the standard deviation. One advantage of the fractional difference model is that it has been shown to provide good forecasting performance regardless of the long memory generating mechanism \citep{VeraValdes2020a}.

\subsubsection{Forecasting with cross-sectional aggregation}\label{sec:forecasting_csa}

Another way to obtain the autoregressive coefficients is by solving the Yule-Walker equations for the model. In the case of cross-sectional aggregation, the Yule-Walker equations are given by:
\begin{equation}
    \begin{bmatrix}	\psi_1 \\ \psi_2 \\ \vdots \\ \psi_k	\end{bmatrix} = \begin{bmatrix}	1 &\rho_{CSA}(1) &\cdots &\rho_{CSA}(k-1) \\\rho_{CSA}(1) & 1 &\cdots &\rho_{CSA}(k-2)\\ \vdots &\vdots &\ddots &\vdots \\ \rho_{CSA}(p-1)&\rho_{CSA}(k-2) &\cdots &1\end{bmatrix}^{-1}
    \begin{bmatrix}	\rho_{CSA}(1) \\\rho_{CSA}(2) \\ \vdots \\\rho_{CSA}(k) \end{bmatrix},
\end{equation}
with $\rho_{CSA}(\cdot)$ defined as in (\ref{ACF:CSA}).

\LMjl\ solves the system above and produces the forecast in the \fct{csa\_forecast} function. Similarly to the fractional difference model, the function \fct{csa\_forecast\_plot} produces the plot. The confidence bands are also available by specifying the standard deviation.

\subsubsection{Forecasting the HAR model}\label{sec:forecasting_har}

A final way to obtain the autoregressive parameters is when the model is already estimated in an autoregressive form, as is the case for the HAR model. Therefore, forecasting using the HAR model is straightforward. We should only iteratively construct the regressors given the restrictions imposed by the model.

\LMjl\ produces the forecast for the HAR model via the function \fct{har\_forecast}, while the plot is produced using the function \fct{har\_forecast\_plot}. The functions estimate the HAR model as part of the forecast given the recursive nature of the model. Hence, custom HAR models are forecasted by specifying the lags to include in the model using the optional input array \code{m}.

\subsection{Forecasting illustration in Nile River minima}\label{sec:forecasting_nile}

\begin{figure}[ht!]
    \centering
    \includegraphics[width=0.9\columnwidth]{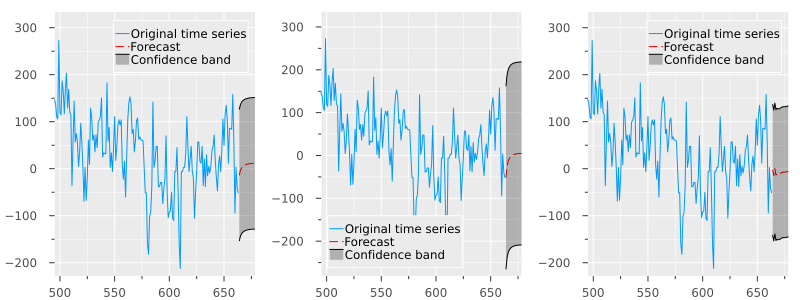}
    \caption[]{Nile River minima forecasts using the fractional difference model (left), cross-sectional aggregation model (centre) and the HAR model (right).}\label{fig:NileForecast}
\end{figure}

As an illustration, Figure~\ref{fig:NileForecast} shows the forecast for the time series of the Nile River minima using the three models through the functions \fct{fi\_forecast\_plot}, \fct{csa\_forecast\_plot}, and \fct{har\_forecast\_plot}. The forecasts are made for the next 30 observations. To make comparisons easier, the forecasts are done in the de-meaned series to avoid the uncertainty regarding the estimation of the mean between models. Furthermore, the figure shows the 95\% confidence intervals. The confidence intervals are computed using the variance of the error term estimated by the models.

The following lines of code show how to generate Figure~\ref{fig:NileForecast}. The code first estimates the models, retrieves the parameters, and passes them on to the forecasting functions. The forecast is made for the next 30 observations.
\begin{CodeChunk}
    \begin{CodeInput}
    using LongMemory, Plots

    Data = NileData()
    dmle, sigfi = fi_mle_est( Data.NileMin )
    pmle, qmle, sigcsa = csa_mle_est( Data.NileMin )
    p1 = fi_forecast_plot( Data.NileMin , 30, dmle, sigfi )
    p2 = csa_forecast_plot( Data.NileMin , 30, pmle, qmle, sigcsa )
    p3 = har_forecast_plot( Data.NileMin , 30; m = [1, 7] )
    l = @layout [a b c]
    theme(:ggplot2)
    plot(p1, p2, p3, layout = l, xlabel = "", ylabel = "" )
    xlims!(500, 673)
    \end{CodeInput}
\end{CodeChunk}


\section{Package usage and benchmarks}\label{sec:software} 

\subsection{Installation and usage}

The package is listed in the \Julia\ general registry, so the installation can be done by the following lines of code.
\begin{CodeChunk}
    \begin{CodeInput}
    julia> using Pkg
    julia> Pkg.add("LongMemory")
    \end{CodeInput}
\end{CodeChunk}

Once installed, the package can be loaded using the following line of code.
\begin{CodeChunk}
    \begin{CodeInput}
    julia> using LongMemory
    \end{CodeInput}
\end{CodeChunk}

The package includes data examples from some of the most common long memory time series in the literature. The Nile River minima data used in this article are accessible using the function \fct{NileData}, while the Northern Hemisphere temperature data \citep{NHTemp} are available using the function \fct{NHTempData}. In addition, the package includes the functions \fct{NileDataPlot} and \fct{NHTempDataPlot} to produce figures with the data, its autocorrelation, the log-periodogram, and the log-variance plots. Similar figures can be obtained for arbitrary data using the \fct{LMPlot} function.

The package's website, available at \url{https://everval.github.io/LongMemory.jl/}, contains all the documentation, including examples for all functions. Moreover, accompanying \proglang{Jupyter} notebooks containing all of the code in this paper are also available at the website.

\subsection{Benchmarks}\label{sec:benchmarks}

This section presents benchmarks contrasting different implementations to show the computational efficiency of \LMjl. The code for the benchmarks is available on the package's website. The benchmarks were made using the \fct{BenchmarkTools} package \citep{BenchmarkToolsJL}.

Table\ref{tab:function_performance} presents the summary of the benchmarks. The following subsections discuss them.

\begin{table}[ht!]
    \centering
    \begin{tabular}{lccr}
        \hline
        Function & Mean & Median & Language:Package \\
        \hline
        \multicolumn{3}{l}{Recursion versus cumulative formulation} \\
        \fct{fi\_rec}$^*$ & 2.063E+04 & 1.320E+04 & \Julia \\
        \fct{fi\_cum}$^*$ & 4.023E+04 & 3.080E+04 & \Julia \\
        \hline
        \multicolumn{4}{l}{Cross-sectional aggregation finite versus asymptotic formulation} \\
        \fct{csa\_gen} & 9.521E+05 & 8.298E+05 & \Julia:\LMjl \\
        \fct{csa\_gen\_fin}$^\dag$ & 1.310E+09 & 1.339E+09 & \Julia:\LMjl \\
        \hline
        \multicolumn{2}{l}{Long memory generation}          \\
        \fct{fi\_gen} & 7.048E+05 & 5.812E+05 & \Julia:\LMjl \\
        \fct{fracdiff.sim} & 1.017E+08 & 1.010E+08 & \R:\pkg{fracdiff} \\
        \hline
        \multicolumn{2}{l}{Fractional differencing}    \\
        \fct{fracdiff} & 7.101E+05 & 5.824E+05 & \Julia:\LMjl \\
        \fct{diffseries} & 2.124E+06 & 1.892E+06 & \R:\pkg{fracdiff} \\
        \fct{fdiff} & 4.150E+06 & 3.950E+06 & \R:\pkg{LongMemoryTS} \\
        \hline
        \multicolumn{2}{l}{Long memory estimation}                      \\
        \fct{gph\_est} & 4.165E+04 & 3.330E+04 & \Julia:\LMjl \\
        \fct{gph} & 1.662E+05 & 1.471E+05 & \R:\pkg{LongMemoryTS} \\
        \fct{fdGPH} & 7.058E+06 & 6.022E+06 & \R:\pkg{fracdiff} \\
        \hline
    \end{tabular}
    \caption{Comparison of function performance. All sample sizes are $10^4$.\\ $^*$These functions are not part of \LMjl. Recursion is implemented as part of the generation routine. \\ $^\dag$\fct{csa\_gen\_fin} stands for the finite sample version of the function \fct{csa\_gen}, both versions are called using the same command,  \fct{csa\_gen}, using multiple dispatch; see Section~\ref{sec:csa}.}
    \label{tab:function_performance}
\end{table}

\subsubsection{Recursive versus non-recursive implementation}\label{sec:benchmarks_recursive}

Fractional differencing is a computationally intensive model. The model requires the computation of the gamma function, for which no closed-form expression in terms of elementary functions exists for arbitrary values. Moreover, the gamma function extends the factorial function so that it grows faster than the exponential function. Already the gamma evaluated at $172$ is larger than the largest \code{Float64} number: $1.7976931348623157\times10^{308}$. Therefore, alternative implementations are needed to reduce the computational burden.

Instead of evaluating the gamma function, we use the recursive implementation based on the identity: $\Gamma(x+1) = x\Gamma(x)$. Recursion can be implemented using a \code{for} loop or the \Julia\ function \fct{cumprod} that computes the cumulative product. The benchmarks for both formulations, presented in the first two rows of Table~\ref{tab:function_performance}, show that the implementation of the \code{for} loop is significantly faster and therefore is the one used in \LMjl.

\subsubsection{Finite versus asymptotic formulation}\label{sec:benchmarks_csa}

Referring to the cross-sectional aggregation model, Table~\ref{tab:function_performance} shows the benchmarks for the finite sample and asymptotic versions of the generation function. The table shows that the finite sample version is significantly slower than the asymptotic version. For a sample size of $10^4$, the finite sample version is at least $10^3$ times slower than the asymptotic version. 

The large difference in performance can be explained by the fact that the finite sample version requires the generation of many individual autoregressive processes (\ref{eq:csa_ind_ar}), while the asymptotic version only requires the computation of the limiting moving average representation (\ref{eq:csa_ma}). 

Note that the asymptotic version achieves similar performance as the long memory generation function by fractional differencing. The latter is a direct consequence of the fact that the asymptotic version of cross-sectional aggregation relies on an analogous use of the fast Fourier transform as the one used for fractional differencing.

As argued in Section \ref{sec:csa}, the asymptotic version is therefore recommended for most applications. The finite sample version can be useful for small sample sizes and for illustrative purposes.

\subsubsection{Comparing speed against other software}\label{sec:comparison}

We compare the performance of \LMjl\ against the packages \pkg{fracdiff} and \pkg{LongMemoryTS} for \proglang{R}. The package \pkg{fracdiff} is the most popular package for long memory time series in \proglang{R}, measured by the number of CRAN downloads. \pkg{LongMemoryTS} is the package that is closer to \LMjl\ in terms of functionality.

For long memory generation, Table~\ref{tab:function_performance} shows the benchmarks for the function \fct{fi\_gen} in \LMjl\ and the function \fct{fracdiff.sim} in the package \pkg{fracdiff}. \pkg{LongMemoryTS} does not provide a function to directly generate processes with long memory. The results show that \LMjl\ is more than $10^2$ times faster than \pkg{fracdiff} at this sample size.

Regarding fractional differencing, Table~\ref{tab:function_performance} shows the benchmarks for the function \fct{fracdiff} in \LMjl\ and the functions \fct{diffseries} and \fct{fdiff} in packages \pkg{fracdiff} and \pkg{LongMemoryTS}, respectively. All functions use the fast algorithm developed by \cite{Jensen2014}. Note that \LMjl\ is the fastest implementation by a large margin. 

Finally, for long memory estimation, Table~\ref{tab:function_performance} shows the benchmarks for the function \fct{gph\_est} in \LMjl and the functions \fct{fdGPH} and \fct{gph} in packages \pkg{fracdiff} and \pkg{LongMemoryTS}, respectively. The benchmarks show that \LMjl\ is significantly faster, taking advantage of the speed of \Julia.


\section{Conclusion} \label{sec:conclusion}

This paper introduces \LMjl, a \Julia\ package for long memory time series. The package includes functions to generate, estimate, and forecast long memory time series. \LMjl\ is the first package publicly available for long memory modelling in \Julia. For some of the methods, \LMjl\ presents the first publicly available implementations in any programming language.

Long memory generation methods include fractional differencing, stochastic duration shocks, and cross-sectional aggregation in asymptotic and finite sample versions. Several estimators are considered, including parametric and those based on log-periodogram regression. Forecasting functions are included for all parametric estimators. We show that \LMjl\ is fast compared to current alternatives. The package also includes data examples from some of the most common long memory time series in the literature and provides plotting capabilities for long memory analysis.


\section*{Computational details}

The results in this article were obtained using \proglang{Julia}~1.10 with the \pkg{LongMemory.jl}~0.1.1 package. \proglang{Julia} and all the packages used are available from the General Registry at \url{https://github.com/JuliaRegistries/General}.

The included data are deployed using \pkg{Artifacts}, hence the package requires \Julia\ 1.6 or later.

Comparisons against other software were made using \proglang{R}~4.3.1 with the \pkg{fracdiff}~1.5-2 and \pkg{LongMemoryTS}~0.1.0 packages. To simplify the comparisons, the benchmarks presented in this article were made using \pkg{BenchmarkTools.jl} through \pkg{RCall.jl} \citep{RCallJL}. Benchmarks using \pkg{microbenchmark} \citep{R:microbenchmark} in \proglang{R} show similar results, as shown in the accompanying \proglang{Jupyter} notebook.

Notebooks with all the codes in this paper and additional examples are available at \url{https://everval.github.io/LongMemory.jl/}. 



\bibliography{refs}






\end{document}